\documentstyle[preprint,aps]{revtex}
\tightenlines
\preprint{\vbox{  
\hbox{IFT-P.016/98}   
\hbox{hep-ph/9802311}
\hbox{February 1998} }}  
\begin{document}  
\draft 
\title{
Left-right asymmetries in polarized lepton-lepton scattering}
\author{J. C. Montero, V. Pleitez and M. C. Rodriguez} 
\address{
Instituto de F\'\i sica  Te\'orica\\ 
Universidade  Estadual Paulista\\
Rua Pamplona, 145\\ 
01405-900-- S\~ao Paulo, SP\\ 
Brazil } 
\date{today}
\maketitle 
\begin{abstract}  
Several parity violating left-right asymmetries in M\o ller electron-electron 
and muon-muon scattering are considered in the context of the electroweak 
standard model at tree level in fixed target and collider experiences. We show 
that in colliders the asymmetry with only one of the beam polarized is large 
enough to compensate the smaller cross section at high energies. We also show 
that these asymmetries are very sensitive to a doubly charged vector bilepton 
resonance but they are insensitive to scalar ones.
\end{abstract}
\pacs{PACS   numbers: 13.88.+e; 
12.60.-i 
12.60.Cn;  
}
\narrowtext   
  
\section{Introduction}
\label{sec:intro}  
It is well known that the so called Next Linear Collider (NLC) will provide
opportunities for both discovery and precision measurements~\cite{nlc}.
With the construction of the next generation of $e^+e^-$ linear 
colliders, with a center of mass energy up to 1500 GeV and which will 
be able to operate also in $\gamma\gamma$, $\gamma e^-$ and $e^-e^-$ modes, new 
perspectives arise in detecting new physics beyond the electroweak standard 
model in processes having non-zero initial electric charge (and non-zero lepton 
number). For instance, new resonances of doubly charged scalar and vector 
bosons will enhance the cross section and in this case, it is possible that the 
measurement of the left-right asymmetry ($A_{RL}$) in electron-electron 
scattering can be a better parameter to signal out those resonances (if any) 
than the total cross section itself. On the other hand, although the First Muon 
Collider (FMC) will be one of the type $\mu^-\mu^+$ (it is a circular 
machine)~\cite{gunion}, if this sort of technology is dominated perhaps a Linear 
Muon Collider (LMC) will be feasible.
In this work we will consider mainly $e^-e^-$ processes but we will also 
comment the muon case.

\subsection{Lepton M\o ller scattering at high energies}
\label{subsec:moller}
The appealing features for studying the parity violating asymmetries between
the scattering of left- and right-handed polarized electrons on a variety of 
targets was pointed out some years ago by Derman and Marciano~\cite{dm}. 
In particular, in the context of the electroweak standard model (ESM) the 
measurement of the $A_{RL}$ asymmetry  determined the relative 
sign between the weak and electromagnetic interactions. 

It is well known that the left-right asymmetry in electron-electron M\o ller 
scattering (fixed target) and unpolarized target is rather small 
$\sim10^{-7}$ (see below)~\cite{dm}. 
Notwithstanding, since the scattering has a large cross section, we 
can expect that fixed target experiences like those at SLAC~\cite{slac} will
have enough number of events to determine such a small number. The typical beam
energy of such experiences is 50 GeV and both the incident electron beam and 
the target electron can be polarized. 
On the other hand, collider experiences were stopped years ago at very low 
energies, around 1.2 GeV~\cite{co1}. 
Although the $A_{RL}$ asymmetry is larger in colliders 
than in fixed target experiments, the former have a 
smaller cross section. Hence, it seems at first sight that it will be 
difficult a measurement of that asymmetry with large statistics in 
colliders. 

In general we have processes like 
\begin{equation}
l_1(p_1,\lambda_1) + l_2(p_2,\lambda_2)\to l'_1(q_1,\Lambda_1) + 
l'_2(p'_2,\Lambda_2)
\label{ll}
\end{equation}
when $l_{1,2}=e,\mu$ and $l'_{1,2}=e,\mu,\tau,L$, with $L$ an exotic
lepton. In Eq.~(\ref{ll}) $p_i,q_i$ ($\lambda_i,\Lambda_i$) denote 
the initial and final lepton momenta (helicities). In this work we 
will consider only processes like
\begin{equation}
l(p_1,\lambda_1) + l(p_2,\lambda_2)\to l(q_1,\Lambda_1) + 
l(q_2,\Lambda_2)
\label{ll1}
\end{equation}
Note that if we were considering quarks in the final state, there must be a 
mixing of ordinary quarks with exotic ones since with the known quarks we can 
not build a doubly charged state. 

Our main goal is to compare the left-right asymmetry in the context of the
electroweak standard model in fixed target experiments with the respective one in colliders
experiments. Since we are not neglecting the masses of the fermions we
can use the same amplitudes for the case of $e^-e^-$ and $\mu^-\mu^-$
scattering. We also study the behaviour of that quantity when a vector or
scalar bilepton is considered in collider experiences.

\subsection{4-spinor helicity eigenstates and left-right asymmetry}
\label{subsec:sub1}

In this work we will not neglect the ESM Higgs-lepton 
interactions because in collider 
experiences with muons the Higgs contributions are not, in principle, 
negligible.  
Beside, in the future we will consider processes in which the 
final lepton can be $\tau$ or even an exotic 
lepton $L^-$ (or even exotic quarks $J$). 
Hence, we will work with massive 4-spinor eigenvectors of the helicity 
operator, defined as
\begin{equation}
\Sigma=
\left(
\begin{array}{cc}
\vec{\sigma}\cdot\vec{p}/\vert\vec{p}\vert & 0 \\
0 &\vec{\sigma}\cdot\vec{p}/\vert\vec{p}\vert
\end{array}
\right),
\label{helicity}
\end{equation}
with the gamma matrices in the chiral representation, that are given by:
 
\begin{equation}
u_{lR}=N_l\,\left(
\begin{array}{c}
(1+\eta_l)\cos(\theta_l/2)\,e^{-i\phi_l/2} \\
(1+\eta_l)\sin(\theta_l/2)\,e^{i\phi_l/2} \\
(-1+\eta_l)\cos(\theta_l/2)\,e^{-i\phi_l/2}\\
(-1+\eta_l)\sin(\theta_l/2)\,e^{i\phi_l/2}
\end{array}
\right),\quad u_{lL}=N_l\,\left(
\begin{array}{c}
(-1+\eta_l)\sin(\theta_l/2)\,e^{-i\phi_l/2} \\
-(-1+\eta_l)\cos(\theta_l/2)\,e^{i\phi_l/2} \\
(1+\eta_l)\sin(\theta_l/2)\,e^{-i\phi_l/2}\\
-(1+\eta_l)\cos(\theta_l/2)\,e^{i\phi_l/2}
\end{array}
\right),
\label{e1a}
\end{equation}

where we have defined 
\begin{equation}
N_l=\sqrt{\frac{E_l}{(1+\eta_l^2)}}, \quad \eta_l^2=\frac{E_l+m_l}{E_l-m_l},
\label{e2}
\end{equation}
being $E_l$ and $m_l$ the total energy and rest mass for the lepton $l$. 
($\hbar=1$ and $c=1$ throughout this work.) In Eq.~(\ref{e1a}) 
$u_R,u_L$ denote positive energy spinors
with positive (right) and negative (left) helicity, respectively. 
Similarly expressions can be obtained for the negative energy solutions but we 
will not write them explicitly.

The left-right asymmetry is defined as
\begin{equation}
A_{RL}(ll\to ll)=\frac{d\sigma_R-d\sigma_L}{d\sigma_R+d\sigma_L},
\label{asy1}
\end{equation}
where $d\sigma_{R(L)}$ is the differential cross section for one right
(left)-handed lepton $l$ scattering on an unpolarized lepton $l$.
That is
\begin{equation}
A_{RL}(ll\to ll)=\frac{(d\sigma_{RR}+d\sigma_{RL})-(d\sigma_{LL}
+d\sigma_{LR})}{(d\sigma_{RR}+d\sigma_{RL})+(d\sigma_{LL}
+d\sigma_{LR})},
\label{asy2}
\end{equation}
where $d\sigma_{ij}$ denotes the cross section for incoming leptons with
helicity $i$ and $j$, respectively, and they are given by 
\begin{equation}
d\sigma_{ij}
\propto\sum_{kl}\vert M_{ij;kl}\vert^2,\quad i,j;k,l=L,R.
\label{dsigma}
\end{equation}

When both leptons are  
identical we can use the property $M_{RL;RL}=M_{LR;LR}$ (which implies 
$d\sigma_{RL}=d\sigma_{LR}$) assured by rotational invariance in the former 
case. Hence, for calculating the asymmetry $A_{RL}$ in Eq.~(\ref{asy2}) we
use
\begin{mathletters}
\label{sigmas}
\begin{equation}
d\sigma_{RR}\propto\vert M_{RR;RR}\vert^2+\vert M_{RR;LR}\vert^2+
\vert M_{RR;RL}\vert^2+\vert M_{RR;LL}\vert^2,
\label{s1}
\end{equation}

\begin{equation}
d\sigma_{RL}\propto\vert M_{RL;RR}\vert^2+
\vert M_{RL;RL}\vert^2+\vert M_{RL;LR}\vert^2+
\vert M_{RL;LL}\vert^2,
\label{s2}
\end{equation}

\begin{equation}
d\sigma_{LR}\propto\vert M_{LR;RR}\vert^2+\vert M_{LR;RL}\vert^2+
\vert M_{LR;LR}\vert^2+\vert M_{LR;LL}\vert^2,
\label{s3}
\end{equation}
and
\begin{equation}
d\sigma_{LL}\propto\vert M_{LL;RR}\vert^2+\vert M_{LL;RL}\vert^2+
\vert M_{LL;LR}\vert^2+\vert M_{LL;LL}\vert^2.
\label{s4}
\end{equation}
\end{mathletters}

Another interesting possibility is the case when both leptons are 
polarized. We can define an asymmetry $A_{R;RL}$ in which
one beam is always in the same polarization state, say right-handed, and 
the other  is either
right- or left-handed polarized (similarly we can define  $A_{L;LR}$):
 \begin{equation}
A_{R;RL}=\frac{d\sigma_{RR}-d\sigma_{RL}}{d\sigma_{RR}+d\sigma_{RL}},\qquad
A_{L;RL}=\frac{d\sigma_{LR}-d\sigma_{LL}}{d\sigma_{LL}+d\sigma_{LR}}.\qquad
\label{ar}
\end{equation}

We can define also an asymmetry when one incident particle is right-
handed and the other is left-handed and the final states are right- and left or
left- and right-handed:
\begin{equation}
A_{RL;RL,LR}=\frac{d\sigma_{RL;RL}-d\sigma_{RL;LR}}{d\sigma_{RL;RL} 
+ d\sigma_{RL;LR}}
\label{ar3}
\end{equation}
or similarly, $A_{LR;RL,LR}$. These asymmetries, in Eqs.(\ref{ar}) and 
(\ref{ar3}), are also dominated by QED 
contributions. However, this will not be the case if a bilepton resonance
does exist at typical energies of the NLC. To show this fact is the goal of 
the next section.
These asymmetries can be calculated for both fixed target (FT) and 
colliders (CO) experiments.

This work is organized as following. In Sec.~\ref{sec:sm} we study the
left-right asymmetries for both fixed target and collider experiences
in the context of the ESM and for $e^-e^-\to e^-e^-$. We also briefly consider
the case of $\mu^-\mu^-$ colliders. The same asymmetries for colliders
are shown in Sec.~\ref{sec:331} but considering the contributions of 
vector and scalar bileptons. Our conclusions appear in the last section
while we left the appendices \ref{sec:aa}--\ref{sec:higgs} for showing the 
several amplitudes and cross sections we have used in this work.

\section{Left-right asymmetries in the electroweak standard model}
\label{sec:sm}
Let us consider the lepton-lepton (diagonal) process in the context of 
the electroweak standard model at tree level. The relevant part of the 
Lagrangian of this model is
\begin{equation}
{\cal L}_F=-\sum_i\,\frac{g\,m_i}{2M_W}\,\bar\psi_i\psi_i\,H^0-
e\sum_iq_i\bar\psi_i\gamma^\mu\psi_iA_\mu-\frac{g}{2\cos\theta_W}\,
\psi_i\gamma^\mu(g^i_V-g^i_A\gamma^5)\psi_iZ_\mu,
\label{su21}
\end{equation}
$\theta_W\equiv\tan^{-1}(g'/g)$ is the weak mixing angle, $e=g\sin\theta_W$
is the positron electric charge with $g$ such that
\begin{equation}
g^2=\frac{8G_F M^2_W}{\sqrt2};\quad {\rm or}\quad 
g^2/\alpha=4\pi\sin^2\theta_W,
\label{gf}
\end{equation}
with $\alpha\approx 1/128$; and the vector and axial neutral couplings are
\begin{equation}
g^i_V\equiv t_{3L}(i)-2q_i\sin^2\theta_W,\quad
g^i_A\equiv t_{3L}(i),
\label{gva}
\end{equation}
where $t_{3L}(i)$ is the weak isospin of the fermion $i$ and $q_i$ is the 
charge of $\psi_i$ in units of $e$. 

\subsection{$e^-e^-$ Fixed target experiences}
\label{subsec:ftsmee}

With the spinors in Eqs.~(\ref{e1a}) we have obtained all the helicity
amplitudes of Ref.~\cite{dm} when $m_e\to 0$; even with the actual 
value of $m_e$ the numerical value for the $A_{RL}(ee)$ asymmetry coincides 
with that obtained in Ref.~\cite{dm} ({\it i.e.}, at tree level) as it 
should be. In fact, neglecting the lepton masses we obtain the asymmetry for 
a fixed target (FT) experience in the ESM context:
\begin{equation}
A^{{\rm FT,ESM}}_{RL}(ll\to ll)\approx-\frac{G_F Q^2}{\sqrt2\pi\alpha}\,
\frac{1-y}{1+y^4+(1-y)^4}\,(1-4\sin^2\theta_W),
\label{assi0}
\end{equation}
where $Q^2=-(p'_1-p_1)^2$, $y=\sin^2(\theta/2)$ and the others constants are the 
well know $\alpha$, $G_F$ and $\sin^2\theta_W$. 
For FT experiences we have $Q^2=-q^2=y(p_1+p'_1)=y(2m^2_{l_1}+~
2m_{l_1}E_{\rm beam})$. In Eq.~(\ref{assi0}) the approximation $m^2_e\ll 
Q^2\ll M^2_Z$ was used. 
Also terms of the order $m_l/E_{\rm beam}$ were 
neglected. This approximation is valid even for $E_{\rm beam}\approx1$ TeV 
(typical energies of the NLC) since in this case $Q^2\approx 0.5\; 
({\rm GeV})^2$.  

We see that for a FT experiment, the $A_{RL}$ asymmetry is small,
beside of the factor $1-~4\sin^2\theta_W$ we have a small $Q^2$~\cite{cm}.
Using the helicity amplitudes given in Appendices~\ref{sec:aa} and 
\ref{sec:ab}, with $E_{\rm beam}=50$ GeV, $\theta\approx90^o$ ($y
=1/2$) and for 100\% beam polarization, we obtain, 
\begin{equation}
A^{{\rm FT,ESM}}_{RL}(ee\to ee)\approx-3\times 10^{-7},
\label{aft1}
\end{equation}

Hence, for FT experiments we have confirmed the value obtained in Ref.~\cite{dm} for 
the case of $e^-e^-$ at tree level:
\begin{equation}
A^{\rm FT,ESM}_{RL}(ee\to ee)\approx -6\times 10^{-9}
\left( E_{\rm beam}/1\,{\rm GeV}\right).
\label{aft2}
\end{equation}
Radiative corrections reduce the tree level value 
for the $A^{\rm FT,ESM}_{RL}$ asymmetry in $e^-e^-$ by $40\pm 3\%$~\cite{cm}.
Although the asymmetry is small in FT experiments, the cross section of the 
M\o ller scattering is high: $\sigma_T\approx 10^3$ nb. In Eq.(\ref{ac1})  
we show the differential cross section for the fixed target case. 

The $A^{\rm FT,ESM}_{R;RL}$ asymmetry defined in Eq.~(\ref{ar}) 
is shown in Fig.~1 as a function of $\sin^2(\theta/2)$. This 
asymmetry is dominated in the context of the ESM by purely QED 
effects~\cite{bincer,ft1}. However, we will see that it sensible to
purely electroweak effects. In fact, for massless electrons we have 
(see Table I)
\begin{equation}
A^{\rm FT,ESM}_{R;RL}(ee\to ee)=\frac{\frac{1-y^4-(1-y)^4}{y^2(1-y)^2}+
s\,\beta_W (g^2_V+g^2_A+4g_Vg_A)+O(\beta_W^2)}
{\frac{1+y^4+(1-y)^4}{y^2(1-y)^2} +
s\,\beta_W (3g^2_V+g^2_A+4g_Vg_A)+O(\beta_W^2)},
\label{ar2}
\end{equation}
where $\beta_W=g^2/4\pi\alpha M^2_Z\approx5.2\times10^{-4}$. 
Notice however that there are electroweak contributions which do not have
the suppression factor $1-4s^2_W$. Hence, the last effects are greater than
in the case when the target is unpolarized {\it i.e.,} for the case of 
the asymmetry defined in Eq.~(\ref{asy2}). This is due to the fact that the 
contributions proportional to $g^2_A$ do not cancel out. When we
calculate the asymmetry with the target unpolarized the cancelation
occurs and we obtain the expression given in Eq.~\ref{assi0}.
In Fig.~1 it appears the $A_{R;RL}$ asymmetry when all amplitudes in
Appendix~\ref{subsec:foton} are considered.
In Fig.~2 we show the same asymmetry but only considering
the photon amplitudes which are given in Appendix~\ref{subsec:foton}.
We note that, in fact, the electroweak contributions are important as can be 
seen by comparing Figs.~1 and 2.

For theoretical purely QED calculations in the Born approximation see Refs.~\cite{bincer,rusos} and for radiative correction 
see Ref.~\cite{rusos}. For experimental data see Ref.~\cite{ft1}. 
The difference with the theoretical calculations~\cite{bincer,rusos} is due to 
the electroweak effects that we have considered. This indicates that this 
asymmetry is sensitive to massive vector boson exchanges.
Measurements of the QED contribution to the $A_{R;RL}$ asymmetry 
({\it i.e.,} with both the incident beam and target longitudinally polarized) 
in a fixed target experience with a beam energy up to near 20 GeV, are given 
in Ref.~\cite{ft1}.

\subsection{$e^-e^-$ Collider experiences}
\label{subsubsec:coeesm}

Next, let us consider a collider (CO) experiment with the 
interactions given in Eq.~(\ref{su21}). In this case the transferred 
momentum is
\begin{eqnarray}
Q^2&=&-(p_1'-p_1)^2\nonumber \\ &= &
-m^2_{l'_1}-m^2_{l_1}+\frac{s}{2}-2\cos\theta\left[\left(\frac{s}{4}-m^2_{l_1}-
m^2_{l'_{1}}\right)
\left(\frac{s}{4}-2m^2_{l_1}\right)\right]^{\frac{1}{2}},
\nonumber \\ &\to & s\,\sin^2\frac{\theta}{2},\qquad {\rm when}\;
m_{l_i}\to0.
\label{cine}
\end{eqnarray}
Hence, now $Q^2$ is not a small factor. 

As we said before, in the ESM the total cross section is $\approx 1000$ nb in 
FT and $\approx 10^{-3}$ nb in collider experiences at $\sqrt{s}=300$ GeV. 
(See the Eqs.(\ref{ac1}) and (\ref{ac2}) 
for the differential cross section in the context of the electroweak standard 
model for collider and fixed target experiment, respectively.) 
So, the required luminosity, or the parameter to be measured, in colliders must 
be larger than in fixed target experiments. 
According to Czarnecki and Marciano~\cite{cm} the number of scattering events
required for a $10^{-8}$ statistical accuracy in FT is of the order of 
$10^{16}$.
However, since the cross section for M\o ller scattering at FT is large
at low $Q^2$ the above requirement does not intimidate.
On the other hand, in collider experiments the cross section is rather smaller
than that of FT at high energies. However, as in this case the asymmetry is 
$\sim10^{5}$ times  the corresponding value for FT it is 
possible that $10^5$ events/year, which is the number of events 
expected for a luminosity of ${\cal L}\approx10^{33}\,
{\rm cm}^{-2}{\rm s}^{-1}$, might be enough to measure the 
$A_{RL}$ asymmetry in collider experiments. 
In addition, if new physics do really exist at 
the hundred's GeV level the cross section in colliders can serve to detect new 
resonances. If this is indeed the fact, the cross section will have the
typical resonance enhancement(s).
Recall also that the CO measurements 
of $A_{RL}$ in M\o ller scattering unlike the case of FT experiments 
will not have a background on $eN$ scattering.  

Thus, it is worth to consider the contribution to the $A_{RL}$
asymmetry at the energies of next linear colliders (electrons up 
to 1.5 TeV) and also in muon colliders (muons up to 4 TeV). Although large 
polarization implies sacrifice in luminosity at a muon collider~\cite{gunion}, 
new physics (if any) with clear signature could be distinguished even with low 
luminosity. Anyway, the asymmetry depends more on the polarization than on the 
luminosity.

At energies $\sqrt{s}=300$ GeV and $\theta\approx90^o$ 
(in the center of mass frame) the asymmetry is 
\begin{equation}
A^{\rm CO, ESM}_{RL}(ee\to ee)\approx -0.05,
\label{asec}
\end{equation}
Compare with the value for the FT 
experiment given in Eq.~(\ref{aft1}). 
In Fig.~3 we show the $A^{{\rm CO,ESM}}_{RL}$ asymmetry as a 
function of $\sqrt{s}$ and $\theta$. Notice that this asymmetry is a smooth
function of both variables.


We can integrate in the scattering angle and define the asymmetry 
$\overline{A}_{RL}$ as
\begin{equation}
\overline{A}_{RL}(ll\to ll)=\frac{(\int d\sigma_{RR}+\int d\sigma_{RL})-(\int 
d\sigma_{LL}
+\int d\sigma_{LR})}{(\int d\sigma_{RR}+\int d\sigma_{RL})+(\int d\sigma_{LL}
+\int d\sigma_{LR})},
\label{aint}
\end{equation}
where $\int d\sigma_{ij}\equiv \int^{175^o}_{5^o}d\sigma_{ij}$. This integrated
asymmetry for $ee\to ee$, in the electroweak standard model 
($\overline{A}^{\rm CO,ESM}_{RL}(ll\to ll)$), appears in Fig.~4 as function 
of $\sqrt{s}$. We see that this integrated asymmetry varies slowly with
$\sqrt{s}$ and it is almost constant at the value $-0.02$ above
$\sqrt{s}=400$ GeV. 

\subsection{$\mu^-\mu^-$ Experiences}
\label{subsubsec:mumusm}

An hypothetical fixed target $\mu^-\mu^-$ experience at the same 
$E_{\rm beam}$ energy used in Sec.~\ref{subsec:ftsmee} gives 
$A_{RL}(\mu\mu)\approx5.4\times~10^{-5}$ for $\theta\approx90^o$. In this case
$Q^2_{\rm muon}=(m_e/m_\mu)Q^2_{\rm electron}\approx4.2\,{\rm GeV}^2$. 
A more realistic case is the $\mu^-\mu^-$ collider where, for the same 
experimental conditions than in the previous subsection, we obtain 
\begin{equation}
A^{\rm CO,ESM}_{LR}(\mu\mu\to\mu\mu)\approx-0.1436.
\label{asmc}
\end{equation}
Notice the large value for the asymmetry for the muons with respect to the 
electron case given in Eq.~(\ref{asec}).

In the electron case the Higgs contributions are negligible. For the muon case
if we do not take into account the Higgs amplitudes given in Appendix~
\ref{subsec:h0} the asymmetry in Eq.~(\ref{asmc}) has the value of $-0.1437$
showing a slight dependence on the Higgs contributions.
There are also effects of the lepton mass in the $\gamma,Z$ contributions 
to the asymmetry that are the main responsible for the different values in 
Eqs.~(\ref{asec}) and (\ref{asmc}).
Independently of the numerical value of the Higgs contributions, 
the important issue is that all its main contributions (see the amplitudes
given in given in Sec.~\ref{subsec:h0} cancel out in the
$A_{RL}$ asymmetry defined in Eqs.~(\ref{asy2}). In fact, as 
we will see in the next section, even in models with larger scalar-lepton
couplings the extra scalar contributions cancel out.

\section{Left-right asymmetries in bilepton models}
\label{sec:331}

An example of possible new physics are models with a doubly charged vector 
boson~\cite{331}. Then, the charged current interactions in terms of the 
physical basis is given by
\begin{equation}
-\frac{g}{\sqrt2}\left[\bar\nu_L E^{\nu\dagger}_LE^l_Ll_LW^+_\mu+ 
\bar l^c_L\gamma^\mu E^{lT}_RE^\nu_L \nu_L V^+_\mu-
\bar l^c_L E^{lT}_RE^l_L l_L U^{++}_\mu\right]+H.c.,
\label{cc331}
\end{equation}
with $l'_L=E^l_Ll_L,\quad l'_R=E^l_Rl_R,\quad \nu'_L=E^\nu_L\nu_L$,
the primed (unprimed) fields denoting symmetry (mass) eigenstates.
We see from Eq.~(\ref{cc331}) that for massless neutrinos we have no mixing
in the charged current coupled to $W^+_\mu$ but we still have mixing 
in the charged currents coupled to $V^+_\mu$ and $U^{++}_\mu$. 
That is, if neutrinos are massless we can always chose 
$E^{\nu\dagger}_LE^l_L=1$. However,  
the charged currents coupled to $V^+_\mu$ and $U^{++}_\mu$ are not diagonal in 
flavor space and the mixing matrix ${\cal K}=E^{lT}_RE^\nu_L$ has
three angles and three phases. (An arbitrary $3\times 3$ unitary matrix has 
three angles and six phases. In the present case, however, the matrix 
${\cal K}$ is determined entirely by the charged lepton sector, so we 
can rotate only three phases~\cite{liung}).

If the bilepton $U^{++}_\mu$ is not too heavy, say, with a mass of the order of
a few hundred GeVs it is possible that it will appear as a resonance in the next 
linear colliders (NLC).  
Assuming that the matrix ${\cal K}$ is almost diagonal we can neglect the mixing 
effects in processes involving the same initial and final leptons as $l^-l^-\to 
l'^-l'^-$. The helicity amplitudes for the $U$-exchange in the $s$-channel are 
given, at tree level, in the Appendix~\ref{sec:ad}. 
We do not known yet the total width $\Gamma_U$ of the $U^{--}$ bilepton, 
however, we know that at least it has to decay into $l^-l'^-$:
If these were the only channels we have for the width
\begin{equation}
\Gamma(U^{--}\to l^-_il^-_j)=\frac{G_FM^3_U}{6\sqrt2\pi}\,\vert {\cal K}_{ij}
\vert^2,
\label{gammaul}
\end{equation}
and if ${\cal K}_{ij}$ is almost diagonal the main decays will be $U\to e^-e^-,
\mu^-\mu^-,\tau^-\tau^-$ and with $M_U=300$ GeV, we have 
$\Gamma(U\to{\rm leptons})
\approx 36$ GeV. However, in models with such a vector bilepton there are 
several Higgs fields and also exotic quarks. Hence, we should have to take into
account $U^{--}\to {\rm scalars}$ and also $U^{--}\to d_3\bar{J}^{-5/3}, 
\bar{u}_{1}(\bar{u}_2)J^{-4/3}$, with $d_3$ and $u_{1,2}$ denoting symmetry 
eigenstates of the known quarks and $J^Q$ ($\bar{J}^Q$) 
denotes an exotic quark (anti-quark) of electric charge $Q$. 
The left-handed mixing matrices, defined by the relation among the symmetry 
eigenstates (primed fields) with the mass eigenstates (unprimed  fields)
{\it i.e.}, $U'_L=V^u_LU_L$ and $D'_L=V^d_LD_L$, survive in the Lagrangian 
density. Thus, the partial width is of the form (neglecting the masses of the 
quarks $J$)
\begin{equation}
\Gamma(U^{--}\to d_3\bar{J}^{-5/3})=\frac{CG_FM^3_U}{6\pi\sqrt2}\vert
(V^d_L)_{3i}\vert^2,
\label{gammauq}
\end{equation}
and similarly for the other decays into $u$'s quarks. Notice that the mixing
matrix in Eq.~(\ref{gammauq}) is not necessarily equal to the Cabibbo-
Kobayashi-Maskawa mixing matrix which is defined as $V_{\rm CKM}=
V^{u\dagger}_LV^d_L$. The parameters in both
Eqs.~(\ref{gammaul}) and (\ref{gammauq}) are constrained for other processes
but here we will not consider this constraints. A realistic calculation will 
have to take into account the masses of the exotic quarks $J$. At present, 
however, there is no experimental data bounds on the values of these masses.
Of course, there are also trilinear vector interactions 
$U^{--}\to W^-V^-$ that it will occur if $M_V+M_W<M_U$ ($V^-$ is a singly 
charged bilepton present in the model). As a detailed calculation of $\Gamma_U$
is out of the scope of this work, we will consider $\Gamma_U$ as a free
parameter.

In this model the Yukawa couplings between leptons and a sextet of
scalars transforming as $({\bf6},0)$ under $SU(3)_L\otimes U(1)_N$ and which is
of the form $g_{ab}\overline{(\psi_{aiL})^c}\psi_{bjL}S_{ij}$
where $a,b=e,\mu,\tau$; $i,j$ are $SU(3)$ indices and the matrix $g_{ab}$ denotes a
symmetrical matrix. If this is the only lepton-Yukawa interaction in the model,
by using a discrete symmetry we can avoid the coupling
to the triplet $\eta\sim({\bf3},0)$,
there is no flavour changing neutral currents (FCNC) in this sector and 
the lepton-Higgs sextet couplings are proportional to $m_l/v_S$ where $m_l$ is 
the lepton mass and $v_S$ is the vacuum expectation value (VEV) of the neutral
component of the sextet.
Since there 
are already two other scalar triplets which give the appropriate mass to the 
$W^\pm$ and $Z^0$ vector bosons, it is not necessary that $v_S$ be of the 
order of magnitude than the other vacuum expectation values that are present 
in the model.
Hence, $v_S$ may be of the order of a few GeV since the
heavier lepton is the $\tau$ with a mass of 1.777 GeV~\cite{pdg}. 
Notice that since the neutral part of the sextet is a doublet of $SU(2)$
this fields have $\rho\equiv M^2_W/M^2_Z\cos^2\theta_W=1$ at the tree level.
The doubly charged member $H^{++}_1$ is part of a triplet with its
neutral partner having vanishing vacuum expectation value (if neutrinos do not
get Majorana masses). There is also a doubly charged $H^{++}_2$ which is a
singlet of $SU(2)$.
Strictly speaking the VEV which is in control of $\rho$ (and flavor changing 
neutral currents coupled to $Z$) is that of the triplet which break the 
$SU(3)_L\otimes U(1)_N$ symmetry. For an arbitrary high value for this VEV we 
can arbitrarily close up to the ESM predictions. 
At the one-loop level it is possible that a fine-tuning is required but
this deserve a more detailed study. 

From the experimental point of view a lower bound on the mass of doubly charged
scalar is $m_{H^{++}}>45 $ GeV~\cite{h2,h22} but we recall that the 
phenomenology of the $H^{++}$ derives from its couplings. The processes which
have to be considered are $\gamma/Z\to H^{++}H^{--}$ and 
$H^{--}\to l^-l'^-$~\cite{gunion2}. In the model of 
Ref.~\cite{331} the coupling $W^-W^-$ with
doubly charged scalars does not exist (the vertex always includes single 
charged vector bilepton $V^-$, {\it i.e.}, the coupling which really exist is 
with $V^-W^-$). 

Although the bilepton model we consider here are gauge
invariant theories which has its own scalar spectrum we will
consider in this work the bilepton contributions as an extra 
contributions to the electroweak standard model ones due to the fact
that in any case the neutral Higgs contributions are smaller than that
of the doubly charged vector bilepton.  
We denote this 
by using the $\{{\rm ESM}\}$ and $\{{\rm ESM+U}\}$ labels in 
cross sections and asymmetries.
However, as we will show, the doubly charged scalar bilepton contributions,
denoted by $\{{\rm ESM+H}\}$, cancel out in the numerator of the $A_{RL}$
asymmetry. Hence, in practice our calculations can be considered a truly
effect of the bilepton model of Ref.~\cite{331}.

\subsection{$e^-e^-$ High energy colliders}
\label{subsubsec:nlc}

By using the expression in Eqs.(\ref{ac1}) and (\ref{dcsu}) we can compute
the effect of such bilepton fields. In Fig.~5 we show the ratio
$\sigma^{\rm CO,ESM+U}/\sigma^{\rm CO,ESM}$ for the vector bilepton 
case as a function of $\sqrt{s}$ and 
$\Gamma_U$. We observe that this ratio becomes appreciable for $\sqrt{s}$
near the $U$-resonance as expected. In this case we have chosen and arbitrary
(but reasonable) value of $M_U=300$ GeV to illustrate the cross section 
behavior. We also note that this quantity is smoothly dependent on
the $\Gamma_U$ values. 
Since the cross section could be large at the $U$-peak, if the mass of the $U$ 
is lower than the energies of future collider experiences, it is interesting to 
know the value of the $A_{RL}$ asymmetry at these energies. 

Here, we will not show the analytic expression of $A^{\rm CO}_{RL}$, but at 
the same conditions than Eq.~(\ref{asec}) it gives
\begin{equation}
A^{\rm CO,ESM+U}_{RL}(ee\to ee)= -0.099,
\label{asec331}
\end{equation}
and we see that it is larger than the respective value obtained in the 
context of the ESM (see Eq.~(\ref{asec})). 
A more illustrative quantity is the angular-integrated 
asymmetry defined in Eq.~(\ref{aint}) and
denoted here by $\overline{A_{RL}}\,^{\rm CO,ESM+U}(ee\to ee)$.
We show it in Fig.~6 as a function of 
$\sqrt s$ and $\Gamma_U$. 
We can observe that the dependence of $\Gamma_U$
cancels out in $\overline{A_{RL}}\,^{\rm CO,ESM+U}(ee\to ee)$, so that even if 
we do not use a realistic calculation for $\Gamma_U$ based in all possible decay 
channels predicted by the model, the result will be, in fact, independent of it. 
As expect the asymmetry is larger in the $U$-pole. 
In Fig.~7 we quote the quantity
\begin{equation}
 \delta\, \overline{A}_{RL}(ee\to ee)\equiv 
(\overline{A}\,^{\rm CO,ESM+U}_{RL}-\overline{A}\,^{\rm CO,ESM}_{RL})/
\overline{A}\,^{\rm CO,ESM}_{RL},
\label{deltadef}
\end{equation} 
with the $\overline{A}$'s defined in 
Eq.~(\ref{aint}), as a function of $\sqrt{s}$ and 
$\Gamma_U$. As observed above this quantity is independent of $\Gamma_U$.
$\delta \,\overline{A_{RL}}$ is large at the $U$-resonance, however, we would 
like to stress 
that it remains appreciably large even far from the $U$-peak. That particular 
behaviour suggest that this quantity could be the one to be considered in the 
search for new physics, like the bilepton $U^{--}$, in future colliders.

Next let us consider the scalar contribution in the model. The $s$-channel
amplitudes appear in Appendix~\ref{sec:higgs}. In this model the amplitudes
in Appendix.~\ref{subsec:h0} are used by making $v\to v_S$,
and, as we said before, in models with several scalar multiplets it is
possible that one of the neutral scalars does not contribute significantly to 
the $W^-$ and $Z$ masses. It means that the vacuum expectation value (VEV) 
of such a scalar can be of the order of a few GeV. 
In this case there are some scalar contributions
which are proportional to $m_l/v_S$ with $m_l$ denotes the lepton mass and
$v_S$ is a VEV which gives mass for the leptons. 

Here we will use $M^{++}_{H_i}=50,\, 300$ GeV 
(for both $i=1,2$ although the masses of the triplet and singlet might be 
different) and $v_S=10$ GeV only as an illustration. We also take for
colliders $\sqrt{s}=300$ GeV.
Taken into account the ESM plus the scalar contribution (but not the 
vector $U^{--}$ contributions) we obtain
\begin{equation}
A^{\rm CO,ESM+H}_{RL}(ee\to ee)=-0.05, 
\label{hee}
\end{equation}
for $y=1/2$. We see that this value is the same
of the pure ESM contribution given in Eq.~(\ref{asec}). 
This is not a surprise since 
the purely scalar contributions cancel out in both the neutral Higgs
(as in Appendix.~\ref{subsubsec:mumusm}) and also in the doubly 
charged Higgs contributions as can be seen from the amplitudes in 
Appendix~\ref{sec:higgs} and the definition of the asymmetry in 
Eq.~(\ref{asy2}). However, in the muon-muon case the interference terms
among the neutral Higgs and $\gamma,Z$ contributions are relatively
significant as we will see in the next subsection.

\subsection{$\mu^-\mu^-$ High energy colliders}
\label{subsec:mumuco} 
We can calculate all the above asymmetries in the case of a possible
$\mu^-\mu^-$ collider. For instance taking into account the contributions
of the $U^{--}$ vector bilepton we obtain
\begin{equation}
A^{\rm CO,ESM+U}_{RL}(\mu\mu\to\mu\mu)= -0.1801,
\label{asmcu}
\end{equation}
while the contribution of the $H^{--}$ scalar bilepton gives
\begin{equation}
A^{\rm CO,ESM+H}_{RL}(\mu\mu\to \mu\mu)=-0.1436,
\label{amm}
\end{equation}
for $M_H^{++}=50$ and $300$ GeV 

As discussed above, in the model with bileptons the VEV of the neutral scalar
coupled mainly with leptons does not need to be equal to the VEV of the
ESM ($v\sim246$ GeV) but it can have a lower value, say $v_S\sim10$ GeV.
For the muon case we calculate the asymmetry in Eq.~(\ref{asmcu}) with and
without the ESM Higgs scalar contributions given in Appendix~\ref{subsec:h0}. 
The respective values are $-0.1436$ and $-0.1442$. 
If the mass of the $U^{--}$ were 500 GeV the asymmetry in Eq.~(\ref{asmcu})
has the value $-0.0133$.

\section{Conclusions}
\label{sec:con}

We have considered in this work 
that collider ex\-pe\-ri\-ments could be suitable for studying the left-right
asymmetries in lepton-lepton scattering.
In fact, we have shown that in this case the asymmetries are larger than those
which appear in fixed target experiments. 
In particular we have shown that 
there are significant difference between electron and muon asymmetries, being 
the muon ones larger than the electron ones due to mass effects. However, more 
details were given here for the electron case.
For instance, the $A_{RL}$ asymmetry, when the contributions of the $U^{--}$ 
vector bilepton are considered, is larger than the ESM value, as can be seen 
from Fig.~7 for the electron-electron case.  
On the other hand, the purely scalar bilepton contributions cancel out in the
asymmetries as can be seen from the expression given in 
Appendices~\ref{subsec:h0} and \ref{sec:higgs}. Only the interference terms
between the scalars and $\gamma,Z$ amplitudes survive and these are not
negligible in the muon-muon case.

Concerning the asymmetry $A_{R;RL}$ defined in 
Eq.~(\ref{ar}) {\it i.e.,} when both beams are polarized, we see from Table II 
that it is given by
\begin{equation}
A^{\rm CO,ESM+U}_{R;RL}(ee\to ee)\approx -1+O(\beta_W),
\label{ar4}
\end{equation}
and comparing with Eq.~(\ref{ar2}) we see that the maximal value of
$A^{\rm CO,ESM+U}_{R;RL}(ee\to ee)$ and its dependence on $y$ are different in 
models with vector bileptons and the electroweak standard model 
(almost purely QED) case.
At the $U$-resonance $d\sigma_{RL}\gg d\sigma_{RR}$, hence $A^{{\rm CO},
ESM+U}_{R;RL}(ee\to ee)\approx-1$. In the ESM by the contrary we have 
$d\sigma_{RR}\gg d\sigma_{RL}$, hence $A_{R;RL}^{\rm CO,ESM}(ee\to ee)
\approx+1$. Similarly, $A^{\rm CO,ESM+U}_{L;RL}(ee\to ee)\approx-1$ and 
$A^{\rm CO,ESM}_{L;RL}(ee\to ee)\approx+1$.
Although this asymmetry is large in both the ESM (see Fig.~1) and the bilepton 
model the sign of it is opposite and could be useful for discovering such a
vector field. 

Summarizing, we have calculated in this work several left-right asymmetries in 
the context of the ESM and in models with both vector and scalar bileptons.

Finally a remark. In some models with vector bileptons like $U^{--}$ the decay 
$\mu^-\to e^-e^-e^+$ (and similar ones) does occur. The branching ratio for 
this decay is $\sim10^{-12}$~\cite{pdg}.
This bound strongly limits the ${\cal K}_{e\mu}$ couplings. The branching 
fraction for $\mu\to 3e$ decay is
\begin{equation}
B(\mu\to 3e)\propto \left(\frac{{\cal K}_{e\mu}{\cal K}_{ee}}{M^2_U}\right)^2\,
\frac{1}{G_F^2}.
\label{mu3e}
\end{equation}
For $M_U=300$ GeV and ${\cal K}_{ee}\approx1$ the experimental value for the
above branching ratio implies ${\cal K}_{e\mu}\sim10^{-6}$. 
The doubly charged scalar will also contribute to these rare decays.

\acknowledgments 
This research was partially supported by Conselho Nacional de 
Ci\^encia e Tecnologia (J.C.M. and V.P.) and fully financed  by Funda\c c\~ao 
de Amparo \`a Pesquisa do Estado de S\~ao Paulo (M. C. R.).

\appendix
\section{Invariant amplitudes in the ESM}
\label{sec:aa}

In this appendix we give the invariant amplitudes for the M\o ller scattering
in the context of the $SU(2)_L\otimes U(1)_Y$ (ESM) model. Although the Higgs 
boson in context of the ESM gives contributions that are negligible for practical
proposes, we include its contribution. In models with a rich scalar
spectrum and/or with exotic heavy fermions the scalar contribution may become 
important. We omit a common factor $(2E_l)^2$ in all amplitudes since it 
cancels out in the asymmetries for diagonal processes like $e^-e^-\to e^-e^-$.

\subsection{Photon amplitudes}
\label{subsec:foton}
Up to a factor $ie^2/t$ the $t$-channel amplitudes are (recall $y=
\sin^2(\theta/2)$; $1-2y=\cos\theta$ and $2y^{1/2}(1-2y)=\sin\theta)$
\begin{mathletters}
\label{ft}
\begin{equation}
M^\gamma_{RR;RR}(t)=\frac{4E^2_l+[(1-2y)-3]m^2_l}{2E^2_l},
\;\;
M^\gamma_{RR;LR}(t)=M^\gamma_{RR;RL}(t)=\frac{m_l}{2E_l}\,2y^{1/2}(1-2y),
\label{ft1a}
\end{equation}
\begin{equation}
M^\gamma_{RR;LL}(t)=\frac{m^2_l}{E^2_l}y;
\label{ft1b}
\end{equation}

\begin{equation}
M^\gamma_{LR;RR}(t)=M^\gamma_{LR;LL}(t)=M^\gamma_{RR;LR}(t),
\;\;
M^\gamma_{LR;LR}(t)=\frac{2E^2_l-m^2_{l'}}{E^2_l}\,(1-y),\;\;
M^\gamma_{LR;LR}(t)=M^\gamma_{RR;LL}(t);
\label{ft2}
\end{equation}
\begin{equation}
M^\gamma_{RL;RR}(t)=M^\gamma_{RL;LL}(t)=-M^\gamma_{RR;LR}(t),\;
M^\gamma_{RL;LR}(t)=M^\gamma_{LR;LR}(t),\;M^\gamma_{RL;LL}(t)=
M^\gamma_{RR;LL}(t),\;
\label{ft3}
\end{equation}
\begin{equation}
M^\gamma_{LL;RR}(t)= M^\gamma_{RR;LL}(t),\;-M^\gamma_{LL;LR}(t)=
M^\gamma_{LL;RL}(t)= M^\gamma_{RR;LR};\;
M^\gamma_{LL;LL}(t)=M^\gamma_{RR;RR}(t).
\label{ft4}
\end{equation}
\end{mathletters}
photon $u$-channel, up to a factor $ie^2/u$
\begin{mathletters}
\label{fu}
\begin{equation}
M^\gamma_{RR;RR}(u)=\frac{4E^2_l-[3+(1-2y)]m^2_l}{2E^2_l},\;
M^\gamma_{RR;LR}(u)=-M^\gamma_{RR;RL}(u)=M^\gamma_{RR;LR}(t),\;
\label{tu1a}
\end{equation}

\begin{equation}
M^\gamma_{RR;LL}=\frac{m^2_l}{E^2_l}(1-y);
\label{fu1b}
\end{equation}
\begin{equation}
-M^\gamma_{LR;RR}(u)=M^\gamma_{LR;LL}(u)=M^\gamma_{RR;LR}(t),\;
M^\gamma_{LR;LR}(u)=\frac{m^2_l}{E^2_l}\,y,\;
M^\gamma_{LR;RL}(u)=\frac{2E^2_l-m^2_l}{E^2_l}\,y;
\label{fu2}
\end{equation}
\begin{equation}
M^\gamma_{RL;RR}(u)=-M^\gamma_{RL;LL}=M^\gamma_{RR;LR}(t),\;
M^\gamma_{RL;LR}(u)=M^\gamma_{LR;RL}(u),\;
M^\gamma_{RL;RL}(u)=M^\gamma_{RR;LL}(u);
\label{fu3}
\end{equation}
\begin{equation}
M^\gamma_{LL;RR}(u)=M^\gamma_{RR;LL}(u),\; M^\gamma_{LL;LR}=M^\gamma_{LL;RL}(u)
=M^\gamma_{RR;LR}(t),\; 
M^\gamma_{LL;LL}(u)=M^\gamma_{RR;RR}(u).
\label{fu4}
\end{equation}
\end{mathletters}

\subsection{$Z$ Amplitudes}
\label{subsec:z0}

Up to a factor $ig^2/(t-M^2_Z)$, the $t$-channel $Z$-exchanges amplitudes 
are
\begin{mathletters}
\label{zt1}
\begin{equation}
M^Z_{RR;RR}(t)=\left[ \frac{4E^2_l+[-3+(1-2y)]m^2_l}{2E^2_l}\right]
(g^2_V+g^2_A)-4\left( \frac{E^2_l-m^2_l}{E^2_l}\right)^{\frac{1}{2}}g_Vg_A,
\label{zt1a}
\end{equation}

\begin{equation}
M^Z_{RR;LR}(t)=\frac{m_l}{2E_l}(g^2_V+g^2_A)2y^{1/2}(1-2y),\;\;
M^Z_{RR;RL}(t)=\frac{m_l}{2E_l}(g^2_V-g^2_A)2y^{1/2}(1-2y),
\label{zt1b}
\end{equation}
\begin{equation}
M^Z_{RR;LL}(t)= \frac{m^2_l}{2E_l^2}\left[g^2_V\,y+
g^2_A[3+(1-2y)]\right].
\label{zt1c}
\end{equation}
\end{mathletters}

\begin{mathletters}
\label{zt2}
\begin{equation}
M^Z_{LR;RR}(t)=M^Z_{LR;LL}(t)=M^Z_{RR;RL}(t),
\label{zt2a}
\end{equation}
\begin{equation}
M^Z_{LR;LR}(t)=\left(\frac{2E^2_l-m^2_l}{E^2} \right)(g^2_V-g^2_A)
c^2_{\theta/2},\;
M^Z_{LR;RL}(t)=\frac{m^2_l}{E^2_l}(g^2_V-g^2_A)\,y. 
\label{zt2b}
\end{equation}
\end{mathletters}

\begin{equation}
-M^Z_{RL;RR}(t)=M^Z_{RL;LL}(t)=M^Z_{RR;RL}(t),\; 
M^Z_{RL;LR}(t)=M^Z_{LR;RL}(t),\; M^Z_{RL;RL}=M^Z_{LR;LR}(t).
\label{zt3}
\end{equation}

\begin{equation}
M^Z_{LL;RR}(t)=M^Z_{RR;LL}(t),\;
M^Z_{LL;LR}(t)=M^Z_{LL,RL}(t)=M^Z_{RR;LR}(t),\;
M^Z_{LL;LL}=M^Z_{RR;RR}(t).
\label{zt4}
\end{equation}

Up to a factor $ig^2/(u-M^2_Z)$, the $u$-channel $Z$-exchange amplitudes 
are:
\begin{mathletters}
\label{zu1}

\begin{equation}
M^Z_{RR;RR}(u)\left[\frac{4E^2_l-[3+(1-2y)]m_l^2}{E^2_l}\right](g^2_V+g^2_A)-
4\left( \frac{E^2_l-m^2_l}{E^2_l}\right)^{\frac{1}{2}}g_Vg_A,
\label{zu1a}
\end{equation}

\begin{equation}
M^Z_{RR;LR}(u)=M^Z_{RR;LR}(t),\;M^Z_{RR;RL}(u)=-M^Z_{RR;RL}(t),\;
\label{zu1b}
\end{equation}
\begin{equation}
M^Z_{RR;LL}(u)=\frac{m^2_l}{2E^2_l}\left[ g^2_V(1-y)+g^2_A
[(1-2y)-3]\right]
\label{zu1c}
\end{equation}
\end{mathletters}

\begin{mathletters}
\label{zu2}
\begin{equation}
M^Z_{LR;RR}(u)=M^Z_{LR;LL}(u)=M^Z_{RR;RL}(t),\;\;
M^Z_{LR;LR}(u)=\frac{m^2_l}{E^2_l}(g^2_V-g^2_A)(1-y),
\label{zu2ab}
\end{equation}

\begin{equation}
M^Z_{LR;RL}(u)=\left( \frac{2E^2_l-m^2_l}{E^2_l}\right)(g^2_V-g^2_A)y,
\label{zu2c}
\end{equation}
\end{mathletters}

\begin{equation}
M^Z_{RL;RR}(u)=M^Z_{RL;LL}(u)=M^Z_{RR;RL}(t),\;M^Z_{RL;LR}(u)=M^Z_{LR;RL}(u),\;
M^Z_{RL;RL}(u)=M^Z_{LR;LR}(u).
\label{zu3}
\end{equation}

\begin{mathletters}
\label{zu4}
\begin{equation}
M^Z_{LL;RR}(u)=M^Z_{RR;LL}(u),\;M^Z_{LL;LR}(u)=M^Z_{LL;RL}(u)=M^Z_{RR;RL}(t),
\label{zu4a}
\end{equation}
\begin{equation}
M^Z_{LL;LL}(u)=\left[ \frac{4E^2_l-[3+(1-2y)]m^2_l}{E^2_l}\right](g^2_V+g^2_A)
+4\left( \frac{E^2_l-m^2_l}{E^2_l}\right)^{\frac{1}{2}}g_Vg_A.
\label{zu4b}
\end{equation}
\end{mathletters}

\subsection{$H^0$ Amplitudes}
\label{subsec:h0}
Here $H^0$ denotes a neutral Higgs boson with couplings like the ESM Higgs. 
Up to a common $im^2_l/v^2(t-M^2_H)$ factor, the $t$-channel $H$-exchange
amplitudes are 
\begin{mathletters}
\label{th}
\begin{equation}
M^H_{RR;RR}(t)=-\frac{m^2_l}{E^2_l}\,(1-y),\;
M^H_{RR;LR}(t)=M^H_{RR;RL}=-\frac{m_l}{2E_l}2y^{1/2}(1-2y),\;
M^H_{RR;LL}(t)=y;
\label{th1}
\end{equation}

\begin{equation}
-M^H_{LR;RR}(t)=M^H_{LR;LL}(t)=M^H_{RR;LR}(t);\;
M^H_{LR;LR}(t)=M^H_{RR;RR}(t),\; M^H_{LR;RL}(t)=M^H_{RR;LL}(t);
\label{th2}
\end{equation}

\begin{equation}
-M^H_{RL;RR}(t)=M^H_{RL;LL}(t)=M^H_{RR;LR}(t),\;
M^H_{RL;LR}(t)=M^H_{RR;LL}(t),\; M^H_{LR;RL}(t)=M^H_{RR;RR}(t);
\label{th3}
\end{equation}

\begin{equation}
-M^H_{LL;RR}(t)=M^H_{RR;LL}(t);\;-M^H_{LL;LR}(t)=-M^H_{LL;RL}(t)=M^H_{RR;LR}(t),\;
M^H_{LL;LL}(t)=M^H_{RR;RR}(t)
\label{th4}
\end{equation}
\end{mathletters}

Up to a common $im^2_l/v^2(u-M^2_H)$ factor the $H$ exchange $u$ channel
amplitudes are 
\begin{mathletters}
\label{uh}
\begin{equation}
M^H_{RR;RR}(u)=-\frac{m_l^2}{E^2_l}\,y,\;
-M^H_{RR;LR}(u)=M^H_{RR;RL}= M^H_{RR;LR}(t)\;
M^H_{RR;LL}(u)=1-y;
\label{uh1}
\end{equation}

\begin{equation}
M^H_{LR;RR}(u)=M^H_{LR;LL}(u)=M^H_{RR;LR}(t);\;
M^H_{LR;LR}(u)=M^H_{RR;LL}(u),\; M^H_{LR;RL}(u)=M^H_{RR;RR}(u);
\label{uh2}
\end{equation}

\begin{equation}
M^H_{RL;RR}(u)=-M^H_{RL;LL}(u)=M^H_{RR;LR}(t),\;
-M^H_{RL;LR}(u)=M^H_{RR;RR}(u),\; M^H_{RL;RL}(u)=M^H_{RR;LL}(u);
\label{uh3}
\end{equation}

\begin{equation}
M^H_{LL;RR}(u)=M^H_{RR;LL}(u);\;M^H_{LL;LR}(u)=M^H_{LL;RL}(u)=M^H_{RR;LR}(t),\;
M^H_{LL;LL}(u)=-M^H_{RR;RR}(u)
\label{uh4}
\end{equation}
\end{mathletters}

\section{Total Amplitudes in the ESM}
\label{sec:ab}
We define 
\begin{equation}
M_{ij;kl}=\sum_XM^X_{ij;kl}(u)+
\sum_XM^X_{ij;kl}(t)+\sum_XM^X_{ij;kl}(s)
\label{atotal}
\end{equation}
as the total amplitudes for fixed $ij;kl$ polarization and exchanged particles 
$X=\gamma,Z^0,H^0$ and others. In the context of the 
standard model only $t$ and $u$ channel contribute. Notice also
that this contributions arise only when the lepton-lepton scattering
conserve flavors: $l_1l_2\to l_1l_2$ with $l_1$ being equal or not to $l_2$.

\begin{mathletters}
\label{utotalsm}
\begin{eqnarray}
M_{RR;RR}&=&ie^{2}  \left\{ \left( \frac{4E^{2}-3m^2_l}{2E^{2}} \right) 
\left(\frac{u+t}{tu} \right)+ \frac{m^2_l }{2E^{2}}(1-2y) \left( 
\frac{u-t}{tu} \right) \right.  \nonumber \\
&+&\left. \frac{g^{2}}{e^{2}} \left( \frac{4E^{2}-3 m^2_l}{2E^{2}} 
\right) (g_{V}^{2}+g_{A}^{2})  \left[ \frac{u+t-2M_{Z}^{2}}{(t-M_{Z}^{2})
 (u-M_{Z}^{2})} \right] \right.  \nonumber \\
&-& \left. 4 \frac{g^{2}}{e^{2}} g_{V}g_{A} \sqrt{ 
\frac{E^{2}-m^2_l}{E^{2}}} \ \left[ \frac{u+t-2M_{Z}^2}{(t-M_{Z}^{2}) 
(u-M_{Z}^{2})} \right] \right.  \nonumber \\
&+& \left.  \frac{g^{2}}{e^{2}} \frac{m^2_l}{2E^{2}}(1-2y)(g_{V}
^{2}+g_{A}^{2}) \left[ \frac{u-t}{(t-M_{Z}^{2})(u-M_{Z}^{2})} \right] \right. 
\nonumber \\
&-&\left. \frac{m^4_l}{E^{2}e^{2}v^{2}} \left( \frac{1-y}{t-M_{H}^{2}}+ 
\frac{y}{u-M_{H}^{2}} \right) \right\}. 
\label{sm1} 
\end{eqnarray}

\begin{eqnarray}
M_{RR;LR}&=& ie^{2} \left[ \frac{u+t}{tu}+ \frac{g^{2}}{e^{2}} 
\frac{u-t}{(t-M_{Z}^{2})(u-M_{Z}^{2})}(g_{V}^{2}+g_{A}^{2}) 
+ \frac{m^2_l }{e^{2}v^{2}} \frac{t-u}{(t-M_{H}^{2})
(u-M_{H}^{2})} \right] \nonumber \\ &&\mbox{} \cdot  \frac{m_l}{2E}2y^{1/2}(1-2y). 
\label{sm2} 
\end{eqnarray}

\begin{eqnarray}
M_{RR;RL}&=&ie^{2} \left[ \frac{u-t}{tu}+ \frac{g^{2}}{e^{2}} 
\frac{u-t}{(t-M_{Z}^{2})(u-M_{Z}^{2})}(g_{V}^{2}-g_{A}^{2})  
- \frac{m^2_l}{e^{2}
v^{2}} \frac{t+u-2M_{H}^{2}}{(t-M_{H}^{2})(u-M_{H}^{2})} \right] \nonumber \\ &&\mbox{}
\cdot \frac{m_l}{2E} 2y^{1/2}(1-2y). 
\label{sm3} 
\end{eqnarray}

\begin{eqnarray}
M_{RR;LL}&=&ie^{2}  \left[ \left( \frac{y}{t}+ \frac{1-y}{u} \right) 
\frac{m^2_l}{E^{2}}+ \frac{g^{2}}{e^{2}} \left( 
\frac{g_{V}^{2}y+g_{A}^{2}(4-2y)}{t-M_{Z}^{2}}+ \frac{g_{V}^{2}(1-y)-g_{A}^{2}
(2+2y)}{u-M_{Z}^{2}} \right) \frac{m^2_l}{2E^{2}} \right.  
\nonumber \\
&+& \left. \frac{m^2_l}{e^{2}v^{2}} \left( \frac{y}{t-M_{H}^{2}}+ 
\frac{1-y}{u-M_{H}^{2}} \right) \right]. 
\label{sm4} 
\end{eqnarray}

\begin{eqnarray}
M_{LR;RR}&=&\,\ ie^{2} \left[ \frac{u-t}{tu}+ \frac{g^{2}}{e^{2}} 
\frac{u+t-2M_{Z}^{2}}{(t-M_{Z}^{2} )(u-M_{Z}^{2})}(g_{V}^{2}-g_{A}^{2}) 
+\frac{m^2_l}{e^{2}v^{2}} \frac{u-t}{(t-M_{H}^{2})(u-M_{H}^{2})} 
\right] \nonumber \\ &&\mbox{} \cdot\frac{m_l}{2E}2y^{1/2}(1-2y). 
\label{sm5} 
\end{eqnarray}

\begin{eqnarray}
M_{LR;LR}&=& ie^{2}  \left[ \frac{2}{t}- \frac{m^2_l}{E^{2}} \left( 
\frac{u-t}{tu} \right)+ \frac{g^{2}}{e^{2}} \left[ \frac{2}{t-M_{Z}^{2}}- 
\frac{m^2_l}{E^{2}} \left( \frac{u-t}{(t-M_{Z}^{2})(u-M_{Z}^{2})} \right) 
\right](g_{V}^{2}-g_{A}^{2}) \right.  \nonumber \\
&-& \left. \frac{m^2_l}{e^{2} v^{2}} \left( \frac{m^2_l}{
E^{2}(t-M_{H}^{2})}- \frac{1}{u-M_{H}^{2}} \right) \right] (1-y). 
\label{sm6} 
\end{eqnarray}

\begin{eqnarray}
M_{LR;RL}&=& ie^{2}  \left[ \frac{2}{u}- \frac{m^2_l}{E^{2}} \left( \frac{
u-t}{tu} \right)+ \frac{g^{2}}{e^{2}} \left[ \frac{2}{u-M_{Z}^{2}}- 
\frac{m^2_l}{E^{2}} \left( \frac{t-u}{(t-M_{Z}^{2})(u-M_{Z}^{2})} 
\right) \right](g_{V}^{2}-g_{A}^{2}) \right.  \nonumber \\
&-& \left.  \frac{m^2_l}{e^{2} v^{2}} \left( \frac{m^2_l}{
E^{2}(u-M_{H}^{2})}- \frac{1}{t-M_{H}^{2}} \right) \right]y. 
\label{sm7} 
\end{eqnarray}

\begin{eqnarray}
M_{LR;LL}&=&\,\ ie^{2} \left[ \frac{u+t}{tu}+ \frac{g^{2}}{e^{2}} 
\frac{u+t-2M_{Z}^{2}}{(t-M_{Z}^{2})(u-M_{Z}^{2})}(g_{V}^{2}-g_{A}^{2}) 
- \frac{m^2_l}{e^{2}v^{2}} \frac{t+u-2M_{H}^{2}}{(t-M_{H}^{2})
(u-M_{H}^{2})} \right]\nonumber \\ &&\mbox{} \cdot \frac{m_l}{2E} 2y^{1/2}(1-2y). 
\label{sm8} 
\end{eqnarray}

\begin{eqnarray}
M_{RL;RR}&=&\,\ ie^{2} \left[ \frac{u+t}{tu}+ \frac{g^{2}}{e^{2}} 
\frac{u-t}{(t-M_{Z}^{2})(u-M_{Z}^{2})}(g_{V}^{2}-g_{A}^{2}) 
-  \frac{m^2_l}{e^{2}v^{2}} \frac{t-u}{(t-M_{H}^{2})
(u-M_{H}^{2})} \right] \nonumber \\ &&\mbox{} \cdot\frac{m_l}{2E} 2y^{1/2}(1-2y). 
\label{sm9} 
\end{eqnarray}

\begin{eqnarray}
M_{RL;LR}&=& -ie^{2}  \left[ \frac{-2}{u}+ \frac{m^2_l}{E^{2}} 
\left( \frac{t-u}{tu} \right)- \frac{g^{2}}{e^{2}} 
\left[ \frac{2}{u-M_{Z}^{2}}- 
\frac{m^2_l}{E^{2}} \left( \frac{t-u}{(t-M_{Z}^{2})(u-
M_{Z}^{2})} \right) \right](g_{V}^{2}-g_{A}^{2}) \right.  \nonumber \\
&+& \left.  \frac{m^2_l}{e^{2} v^{2}} \left( \frac{m^2_l}{
E^{2}(u-M_{H}^{2})}+ \frac{1}{t-M_{H}^{2}} \right) \right] y. 
\label{sm10} 
\end{eqnarray}

\begin{eqnarray}
M_{RL;RL}&=& ie^{2}  \left[ \frac{2}{t}- \frac{m^2_l}{E^{2}} 
\left( \frac{
u-t}{tu} \right)+ \frac{g^{2}}{e^{2}} \left[ \frac{2}{t-M_{Z}^{2}}- 
\frac{m^2_l}{E^{2}} \left( \frac{u-t}{(t-M_{Z}^{2})(u-
M_{Z}^{2})} \right) \right](g_{V}^{2}-g_{A}^{2}) \right.  \nonumber \\
&-& \left.  \frac{m^2_l}{e^{2} v^{2}} \left( \frac{m^2_l}{E^{2}(t-M_{H}^{2})}- 
\frac{1}{u-M_{H}^{2}} \right) \right](1-y) 
\label{sm11} 
\end{eqnarray}

\begin{eqnarray}
M_{RL;LL}&=&\,\ ie^{2} \left[ \frac{u-t}{ut}+ \frac{g^{2}}{e^{2}}\, 
\frac{u+t-2M_{Z}^{2}}{(t-M_{Z}^{2})(u-M_{Z}^{2})}(g_{V}^{2}-g_{A}^{2}) 
+ \frac{m^2_l}{e^{2}v^{2}} \,\frac{t-u}{(t-M_{H}^{2})
(u-M_{H}^{2})} \right]\nonumber \\ &&\mbox{} \cdot \frac{m_l}{2E}2y^{1/2}(1-2y). 
\label{sm12} 
\end{eqnarray}

\begin{eqnarray}
M_{LL;RR}&=&ie^{2}  \left[ \left( \frac{y}{t}+ \frac{1-y}{u} \right) 
\frac{m^2_l}{E^{2}}+ \frac{g^{2}}{e^{2}} \left( \frac{g_{V}^{2}y+
g_{A}^{2}(4-2y)}{t-M_{Z}^{2}}+ \frac{g_{V}^{2}(1-y)+g_{A}^{2}(2+2y)}{u-
M_{Z}^{2}} \right) \frac{m^2_l}{2E^{2}} \right.  \nonumber \\
&-& \left. \frac{m^2_l}{e^{2}v^{2}} \left( \frac{y}{t-M_{H}^{2}}- 
\frac{1-y}{u-M_{H}^{2}} \right) \right]. 
\label{sm13} 
\end{eqnarray}

\begin{eqnarray}
M_{LL;LR}&=&\,\ ie^{2} \left[ \frac{t-u}{tu}+ \frac{g^{2}}{e^{2}} 
\frac{u+t-2M_{Z}^{2}}{(t-M_{Z}^{2})(u-M_{Z}^{2})}(g_{V}^{2}-g_{A}^{2}) 
-  \frac{m^2_l}{e^{2}v^{2}} \frac{t-u}{(t-M_{H}^{2})
(u-M_{H}^{2})} \right] \nonumber \\ &&\mbox{} \cdot \frac{m^2_l}{2E} 2y^{1/2}(1-2y). 
\label{sm14} 
\end{eqnarray}

\begin{eqnarray}
M_{LL;RL}&= &ie^{2}  \left[ \frac{u+t}{tu}+ \frac{g^{2}}{e^{2}} 
\frac{u+t-2M_{Z}^{2}}{(t-
M_{Z}^{2})(u-M_{Z}^{2})}(g_{V}^{2}-g_{A}^{2}) 
- \frac{m^2_l}{e^{2}v^{2}} \frac{t-u}{(t-M_{H}^{2})
(u-M_{H}^{2})} \right]\nonumber \\ &&\mbox{} \cdot \frac{m_l}{2E} 2y^{1/2}(1-2y). 
\label{sm15} 
\end{eqnarray}

\begin{eqnarray}
M_{LL;LL}&=&ie^{2} \left[ \left( \frac{4E^{2}-3m^2_l}{2E^{2}} \right) 
\left( \frac{u+t}{tu} \right)+ \frac{m^2_l}{2E^{2}}(1-2y) \left( 
\frac{u-t}{tu} \right) \right.  \nonumber \\
&+&\left.  \frac{g^{2}}{e^{2}} \left( \frac{4E^{2}-3m^2_l}{2E^{2}} 
\right) (g_{V}^{2}+g_{A}^{2})  \left( \frac{u+t-2M_{Z}^{2}}{(t-M_{Z}^{2}) 
(u-M_{Z}^{2})} \right) \right.  \nonumber \\
&+& \left. 4 \frac{g^{2}}{e^{2}} g_{V}g_{A} \sqrt{ 
\frac{E^{2}-m^2_l}{E^{2}}} \left( \frac{u+t-2M_{Z}^{2}}{(t-M_{Z}^{2}) 
(u-M_{Z}^{2})} \right) \right.  \nonumber \\
&+& \left. \frac{g^{2}}{e^{2}} \frac{m^2_l}{2E^{2}}(1-2y)(g_{V}
^{2}+g_{A}^{2}) \left( \frac{u-t}{(t-M_{Z}^{2})(u-M_{Z}^{2})} \right) \right.  
\nonumber \\
&+& \left. \frac{m^4_l}{E^{2}e^{2}v^{2}} \left( \frac{1-y}{t-M_{H}^{2}}
+ \frac{y}{u-M_{H}^{2}} \right) \right]. 
\label{sm16}
\end{eqnarray}
\end{mathletters}

\section{Differential cross section for fixed target and colliders in the ESM}
\label{sec:ac}
In the context of the standard model in a collider experience we have
\begin{eqnarray}
\frac{d\sigma^{\rm ESM}}{d\Omega}\Biggr|_{\rm CO}&=&
\frac{e^4}{128\pi^2s}
\left\{\frac{1+y^4+(1-y)^4}{y^2(1-y)^2}+\frac{2g^2}{e^2}\left[(g_V^2
+g^2_A)\;\frac{1+2a}{y(1-y)(y+a)[1-y+a]}
+(g^2_V-g^2_A)\cdot \right.\right.\nonumber \\ &&\mbox{}
\left.\left.\left( \frac{(1-y)^2}{y(y+a)}+\frac{y^2}{(1-y)[1-y+a]}
\right)\right]+\frac{g^4}{e^4}(g^2_V-g_A)^2
\left( \frac{(1-y)^2}{(y+a)^2}+\frac{y^2}{(1-y+a)^2}\right)\right.
\nonumber \\ &&\mbox{}
\left.+\frac{g^4}{e^4}(g^4_V+6g^2_Vg^2_A+g^4_A)\frac{(1+2a)^2}
{(y+a)^2[1-y+a]}\right\},
\label{ac1}
\end{eqnarray}
where $a\equiv M^2_Z/s$.

For the fixed target experiment we have
\begin{equation}
\frac{d\sigma^{\rm ESM}}{d\Omega}\Biggr|_{\rm FT}\approx
\frac{e^4}{128\pi^2m_ep_e}\,\frac{p_{e'}^2}{(E_e+m_e)p_{e'}-p_eE_{e'}
\cos\theta_{ee'}}\,\left[ \frac{1+y+(1-y)^4}{y^2(1-y)^2}\right],
\label{ac2}
\end{equation}
where we have written only the main contribution. Here $p_e$ ($E_e$) and 
$p_{e'}$ ($E_{e'}$) 
denote the momentum (energy) of the initial and final electron, respectively; 
$\cos\theta_{ee'}$ is the cosine of the angle between the incident and 
one of the final electrons.

\section{$U$ Amplitudes}
\label{sec:ad}
In this appendix we give the invariant amplitudes for the $ll\to l'l'$ 
scattering in the context of a model with doubly charged vector bilepton 
$U^{--}$. In principle, the Higgs contribution are not negligible.
However, for the sake of simplicity we will not take the extra scalar 
contributions into account.

Up to a common $ig^2/2(s-M^2_U)$ factor, the $s$-channel $U^{--}$--exchange 
amplitudes are

\begin{equation}
M_{RR;RR}^{U}(s)= 0,\quad
M_{RR;LR}^{U}(s)= 0,\quad
M_{RR;RL}^{U}(s)= 0,\quad 
M_{RR;LL}^{U}(s)= 0 
,
\label{a}
\end{equation}

\begin{mathletters}
\label{b}
\begin{equation}
M_{LR;RR}^{U}(s)=- \frac{1}{2}\,
\left(\frac{E_{l}-m_{l}}{E_{l}}\cdot
\frac{E^{\prime}_{l}-m^{\prime}_{l}}{E^{\prime}_{l}}\right)^{\frac{1}{2}}
\left[ 1- \left(\frac{E_{l}+m_{l}}{E_{l}-m_{l}}\cdot
\frac{E^{\prime}_{l}+m^{\prime}_{l}}{E^{\prime}_{l}-m^{\prime}_{l}}\right)^
{\frac{1}{2}}\right]2y^{1/2}(1-2y),  
\label{b1}
\end{equation}
\begin{equation}
M_{LR;LR}^{U}(s)= 
\left(\frac{E_{l}-m_{l}}{E_{l}}\cdot
\frac{E^{\prime}_{l}-m^{\prime}_{l}}{E^{\prime}_{l}}\right)^{\frac{1}{2}}
\left[ 1+ \left(\frac{E_{l}+m_{l}}{E_{l}-m_{l}}\cdot
\frac{E^{\prime}_{l}+m^{\prime}_{l}}{E^{\prime}_{l}-m^{\prime}_{l}}
\right)^{\frac{1}{2}}\right](1-y), 
\label{b2}
\end{equation}
\begin{equation}
M_{LR;RL}^{U}(s)=- 
\left(\frac{E_{l}-m_{l}}{E_{l}}\cdot
\frac{E^{\prime}_{l}-m^{\prime}_{l}}{E^{\prime}_{l}}\right)^{\frac{1}{2}}
\left[ 1- \left(\frac{E_{l}+m_{l}}{E_{l}-m_{l}}\cdot
\frac{E^{\prime}_{l}+m^{\prime}_{l}}{E^{\prime}_{l}-m^{\prime}_{l}}
\right)^{\frac{1}{2}}\right]y, 
\label{b3}
\end{equation}
\begin{equation}
M_{LR;LL}^{U}(s)=- \frac{1}{2}\,
\left(\frac{E_{l}-m_{l}}{E_{l}} \cdot
\frac{E^{\prime}_{l}-m^{\prime}_{l}}{E^{\prime}_{l}}\right)^{\frac{1}{2}}
\left[ 1- \left(\frac{E_{l}+m_{l}}{E_{l}-m_{l}}\cdot
\frac{E^{\prime}_{l}+m^{\prime}_{l}}{E^{\prime}_{l}-m^{\prime}_{l}}
\right)^{\frac{1}{2}}\right]2y^{1/2}(1-2y). 
\label{b4}
\end{equation}
\end{mathletters}

\begin{mathletters}
\label{c}
\begin{equation}
M_{RL;RR}^{U}(s)=- \frac{1}{2}\,
\left(\frac{E_{l}-m_{l}}{E_{l}}\cdot
\frac{E^{\prime}_{l}-m^{\prime}_{l}}{E^{\prime}_{l}}\right)^{\frac{1}{2}}
\left[ 1- \left(\frac{E_{l}+m_{l}}{E_{l}-m_{l}}\cdot
\frac{E^{\prime}_{l}+m^{\prime}_{l}}{E^{\prime}_{l}-m^{\prime}_{l}}
\right)^{\frac{1}{2}}\right]2y^{1/2}(1-2y), 
\label{c1}
\end{equation}

\begin{equation}
M_{RL;LR}^{U}(s)= 
\left(\frac{E_{l}-m_{l}}{E_{l}}\cdot
\frac{E^{\prime}_{l}-m^{\prime}_{l}}{E^{\prime}_{l}}\right)^{\frac{1}{2}}
\left[ 1+ \left(\frac{E_{l}+m_{l}}{E_{l}-m_{l}}\cdot
\frac{E^{\prime}_{l}+m^{\prime}_{l}}{E^{\prime}_{l}-m^{\prime}_{l}}
\right)^{\frac{1}{2}}\right]y, 
\label{c2}
\end{equation}

\begin{equation}
M_{RL;RL}^{U}(s)= - 
\left(\frac{E_{l}-m_{l}}{E_{l}}\cdot
\frac{E^{\prime}_{l}-m^{\prime}_{l}}{E^{\prime}_{l}}\right)^{\frac{1}{2}}
\left[ 1+ \left(\frac{E_{l}+m_{l}}{E_{l}-m_{l}}\cdot
\frac{E^{\prime}_{l}+m^{\prime}_{l}}{E^{\prime}_{l}-m^{\prime}_{l}}
\right)^{\frac{1}{2}}\right]y,
\label{c3}
\end{equation}
\begin{equation}
M^{U}_{RL;LL}(s)= \frac{1}{2}
\left(\frac{E_{l}-m_{l}}{E_{l}}\cdot
\frac{E^{\prime}_{l}-m^{\prime}_{l}}{E^{\prime}_{l}}\right)^{\frac{1}{2}}
\left[ 1- \left(\frac{E_{l}+m_{l}}{E_{l}-m_{l}}\cdot
\frac{E^{\prime}_{l}+m^{\prime}_{l}}{E^{\prime}_{l}-m^{\prime}_{l}}
\right)^{\frac{1}{2}}\right]2y^{1/2}(1-2y). 
\label{c4}
\end{equation}
\end{mathletters}

\begin{equation}
M_{LL;RR}^{U}(s)= 0,\quad
M_{LL;LR}^{U}(s)= 0,\quad
M_{LL;RL}^{U}(s)= 0,\quad
M_{LL;LL}^{U}(s)= 0 
.
\label{d}
\end{equation}

\begin{mathletters}
\label{au}
\begin{equation}
M_{RR}^{U}(s)=M_{RR;RR}^{U}(s)+M_{RR;LR}^{U}(s)+M_{RR;RL}^{U}(s)+
M_{RR;LL}^{U}(s)=0,
\label{au1}
\end{equation}

\begin{eqnarray}
M_{RL}^{U}(s)&=&M_{RL;RR}^{U}(s)+M_{RL;LR}^{U}(s)+M_{RL;RL}^{U}(s)+
M_{RL;LL}^{U}(s)\nonumber \\  
&=&\frac{ig^2}{2(s-M^2_U)} \cdot 
\left(\frac{E_l-m_l}{E_l}\cdot
\frac{E^{\prime}_{l}-m^{\prime}_l}{E^{\prime}_l}\right)^{\frac{1}{2}} 
\left[ 1+ \left(\frac{E_l+m_l}{E_l-m_l}\cdot
\frac{ E^{\prime}_l+m^{\prime}_l }{ E^{\prime}_l-m^{\prime}_l} 
\right)^{\frac{1}{2}}\right](1-2y),
\label{au2}
\end{eqnarray}

\begin{eqnarray}
M_{LR}^{U}(s)&=&M_{LR;RR}^{U}(s)+M_{LR;LR}^{U}(s)+M_{LR;RL}^{U}(s)+
M_{LR;LL}^{U}(s)\nonumber \\ 
&=&\frac{ig^2}{2(s-M^2_U)} \cdot 
\left(\frac{E_l-m_l}{E_l}\cdot
\frac{ E^{\prime}_l-m^{\prime}_l }{ E^{\prime}_l}\right)^{\frac{1}{2}}
\left[ 1+\left(
\frac{E_l+m_l}{E_l-m_l}
\cdot
\frac{E^{\prime}_l+m^{\prime}_l }{ E^{\prime}_l-m^{\prime}_l}
\right)^{\frac{1}{2}}
\right](1-2y),
\label{au3}
\end{eqnarray}

\begin{equation}
M_{LL}^{U}(s)=M_{LL;RR}^{U}(s)+M_{LL;LR}^{U}(s)+M_{LL;RL}^{U}(s)+
M_{LL;LL}^{U}(s) =0.
\label{au4}
\end{equation}
\end{mathletters}

For the collider case, if there exist a doubly charged vector boson $U^{--}$
which is not too heavy, it will appear at the energies of the next linear 
colliders (for electron beams up to 1.5 TeV). In this case the cross section
at the $U$-peak will be larger than in the case of the standard model.
Explicitly, the collider cross section in the case of a $U$-exchange besides
the ESM contributions is:
\begin{eqnarray}
\frac{d\sigma^{\rm ESM+U}}{d\Omega}\Biggr|_{\rm CO}&=&
\frac{d\sigma^{\rm ESM}}{d\Omega}\Biggr|_{\rm CO}+\frac{e^4}{128\pi^2s}\,
\left[ \frac{g^2}{4e^4(1-b)^2}((1-y)^2+y^2)
+\frac{g^2}{e^2(1-b)} \cdot\right.\nonumber \\ &&\mbox{}\left.\left( \frac{y^3-(1-y)^3}{y(1-y)}+
\frac{g^2}{e^2}(g^2_V-g^2_A)\frac{y^2(y+a)-(1-y)^2(1-y+a)}
{(y+a)(1-y+a)}\right)\right]
\label{dcsu}
\end{eqnarray}
where $b=M^2_U/s$. 

\section{$H^{++}$ Amplitudes}
\label{sec:higgs}
Here we will consider the amplitudes of the two doubly charged scalars 
$H^{++}_1$ and $H^{++}_2$ which are present in the model of Ref.~\cite{331}. 
The former is part of a triplet of $SU(2)$ and the later one is a singlet of 
$SU(2)$. Hence, the experimental lower bound on their masses will be different.
Up to a common $2m_l^2(m_l-E)^2/v^2_S(s-M^2_{H_i})\;i=1,2$ factor the 
amplitudes are
\begin{equation}
M^{H_i}_{RR;RR}=-M^{H_i}_{RR;LL}=M^{H_i}_{LL;LL}=-M^{H_i}_{LL;RR}=
\frac{(m_l-E)(E+m_l)}{E^2+m_l^2},
\label{mh1}
\end{equation}
with the other amplitudes vanishing.

\narrowtext

\begin{table}
\caption{ Non-vanishing contributions in the ESM when the lepton
mass is zero. Here $\beta_W=g^2/4\pi\alpha M^2_Z$.}
\begin{center}
\begin{tabular}{l|cccc} 
initial$\ddots$final& RR     & RL           & LL  & LR \\ \hline
RR      & $\frac{1}{y(y-1)}+2s\,\beta_W (g_V+g_A)^2$        & 0      & 0   & 0 \\
RL      & 0         &  $\frac{y}{1-y}+sy\,\beta_W (g^2_V-g_A^2)$  & 0  & 
$\frac{1-y}{y}+s\,\beta_W (1-y)(g^2_V-g^2_A)$\\
LR      &  0 &  $\frac{1-y}{y}+s\,\beta_W (1-y)(g_V^2-g_A^2)$    & 0  &  
$\frac{y}{1-y}+sy\,\beta_W (g^2_V-g^2_A)$\\
LL      &  $\frac{y}{1-y}+2s\,\beta_W (g_V+g_A)^2$ & 0 & 0 & 0  \\
\end{tabular}
\end{center}
\end{table}

\begin{table}
\caption{ Non-vanishing contributions of the $U$-vector bosons when the lepton
mass is zero.}
\begin{center}
\begin{tabular}{l|cccc} 
initial$\ddots$ final        & RR     & RL           & LL  & LR \\ \hline
RR      & 0         & 0                            &  0  & 0 \\
RL      & 0         & $-2(1-y)$   & 0  & 
$2y$\\
LR      & 0         & $-2y$   &  0 &
$2(1-y)$  \\
LL      & 0         & 0                            & 0 & 0 \\
\end{tabular}
\end{center}
\end{table}

\newpage

\begin{center}
{\bf Figure Captions}
\end{center}
\vskip .5cm

\noindent {\bf Fig.~1} The asymmetry with both beams polarized defined
in Eq.~(\ref{ar}) in a fixed target experience for the process 
$e^-e^-\to e^-e^-$ as a function of $\sin^2(\theta/2)$, where $\theta$ is
the center of mass scattering angle.\\  

\noindent {\bf Fig.~2} Same as in Fig.~1 but only considering the photon 
contributions given in Appendix~\ref{subsec:foton}.\\

\noindent {\bf Fig.~3} The asymmetry $A^{\rm CO,ESM}_{RL}$ defined in 
Eq.~(\ref{asy2}) for the $e^-e^-\to e^-e^-$ reaction as a function of 
$\sqrt{s}$ and the angle $\theta$.\\

\noindent {\bf Fig.~4}  The angular--integrated asymmetry 
$\overline{A_{RL}}^{\rm CO,ESM}$ defined in 
Eqs.~(\ref{aint}) for the $e^-e^-\to e^-e^-$ reaction as function of $\sqrt{s}$.\\

\noindent {\bf Fig.~5} The ratio $\sigma^{\rm CO,ESM+U}/\sigma^{\rm CO,ESM}$ 
for total cross section obtained from Eqs.~(\ref{dcsu}) which includes
the $s$-channel $U^{--}$ contribution and (\ref{ac2}) for the ESM for
the reaction $e^-e^-\to e^-e^-$ as a function of $\sqrt{s}$ and
$\Gamma_U$. \\

\noindent {\bf Fig.~6} Same as Fig.~4 but taken into account the contributions
of the bilepton $U^{--}$ as a function of $\sqrt{s}$ and $\Gamma_U$.\\

\noindent {\bf Fig.~7} $\delta\,\overline{A_{RL}}$ defined in Eq.~(\ref{deltadef})
as a function of $\Gamma_U$ and $\sqrt{s}$ for $e^-e^-\to e^-e^-$.\\

\begin{references}
\bibitem{note1} Recently the fact that the left-right asymmetry is large
in colliders has been pointed out by 
Czarnecki and Marciano~\cite{cm2}.
\end{references}
\end{document}